\documentstyle[11pt,newpasp,twoside,epsf]{article}
\def\edcomment#1{\iffalse\marginpar{\raggedright\sl#1\/}\else\relax\fi}
\reversemarginpar

\begin{document}
\title{AGN Population at 10$^{-13}$ erg s$^{-1}$ cm$^{-2}$ \\ 
: Results from Optical Identification of {\it ASCA} Surveys}
\author{Masayuki Akiyama}
\affil{Subaru Telescope, NAOJ, 650 North A'ohoku Place,
Hilo, HI, 96720, U.S.A.}

\begin{abstract}
In this paper, results of optical identification
of {\it ASCA} surveys are summarized. 
To understand luminous AGNs in the $z\leq1$ universe,
the {\it ASCA} AGN sample is still better than samples of AGNs from deep 
{\it Chandra} or {\it XMM-Newton} surveys.
Combining the identified sample of AGNs from {\it ASCA}
Large Sky Survey and Medium Sensitivity Survey, 
the sample of hard X-ray selected AGNs have been expanded
up to $\sim$80 AGNs above the flux limit of 
10$^{-13}$ erg s$^{-1}$ cm$^{-2}$ in the 2--10~keV band.
One fifth of the {\it ASCA} AGNs are absorbed narrow-line AGNs. The 
luminosity distribution of the absorbed AGNs 
is limited below 10$^{45}$ erg s$^{-1}$, though there
are 14 non-absorbed broad-line AGNs are 
detected above the luminosity. 
The result suggests lack of absorbed luminous AGNs (deficiency of type 2 QSO).
\end{abstract}

\vspace{-0.5cm}
\section{Introduction}

A hard X-ray selection is one of good selection methods that can sample
AGNs without bias against heavy absorption to the nucleus.
{\it ASCA} is the first satellite that is sensitive enough to detect
AGNs beyond nearby universe in the hard X-ray band.
Many optical spectroscopic-follow-up 
observations have been and being done for X-ray sources detected in {\it ASCA}
surveys that range from wide-area medium sensitivity surveys
(e.g., Della Ceca et al. 2000) to deep surveys with a few pointings 
(e.g., Georgantopoulos et al. 1997).
To understand luminous AGNs in the $z\leq1$ universe,
the {\it ASCA} AGN sample is still better than samples of AGNs from deep 
{\it Chandra} or {\it XMM-Newton} surveys (Figure 1).

Here, the results of optical identification of the {\it ASCA} Large Sky Survey
(hereafter LSS; Ueda et al. 1999a; Akiyama et al. 2000a) 
and the {\it ASCA} Medium Sensitivity Survey (hereafter MSS; 
Ueda et al. 1999b) are summarized.
An advantage of the two surveys is high completeness of the optical identification
(33/34 for {\it ASCA} LSS and 51/55 for {\it ASCA} MSS, so far).
The {\it ASCA} LSS covers 5 degree$^2$ area near the north galactic pole
down to $2\times10^{-13}$ erg$^{-1}$ cm$^{-2}$ s$^{-1}$. 
The optical identification is done for 34 X-ray sources detected by SIS detector
and 30 AGNs are identified.
On the other hand, the {\it ASCA} MSS is a serendipitous survey 
of GIS pointing observations from 1993 to 1996. The original sample consists of 55
X-ray sources with flux more than $3\times10^{-13}$ erg$^{-1}$ cm$^{-2}$ s$^{-1}$ 
in the northern sky selected 
from 15 arcmin radius areas of 176 GIS field of views.
The sample contains 48 AGNs so far.
The optical identification of the {\it ASCA} MSS is now being extended 
to X-ray sources detected in outer area of GIS and in southern sky observations.
Finally the {\it ASCA} MSS sample will contain more than 100 
identified sources. 
 
\section{Deficiency of Type-2 QSOs}

\begin{figure}
\plotfiddle{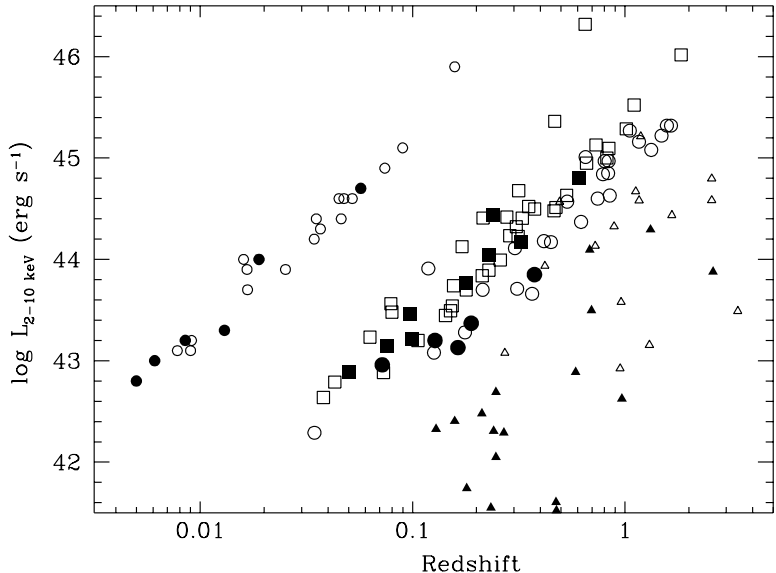}{3.1in}{0}{130}{130}{-430}{-430}
\caption{Redshift-luminosity distributions of hard X-ray selected AGNs.
Small circles, large squares, large circles, and triangles represent
sample of AGNs from the {\it HEAO1} A2, 
{\it ASCA} LSS, {\it ASCA} MSS, and {\it Chandra} Surveys (in detail see text).
Open symbols represent AGNs with X-ray absorption column densities less than
$N_{\rm H}$ of $10^{22}$ cm$^{-2}$ 
 (for {\it ASCA} LSS and {\it HEAO1} A2 samples) and AGNs with a strong broad-line(s)
(for {\it ASCA} MSS and {\it Chandra} samples).}
\end{figure}

A sample of hard X-ray selected AGNs enables us to estimate the 
fraction of absorbed AGNs.
The redshift vs. luminosity distribution 
of the identified AGNs is shown in Figure 1. Squares and large circles
represent {\it ASCA} LSS and MSS AGNs respectively. 
One fifth of identified AGNs
are narrow-line (broad-line-weak) and absorbed AGNs. 
In the luminosity range below $10^{44}$ erg s$^{-1}$,
one third of AGNs are absorbed narrow-line AGNs.
Considering the fact that objects with absorption column densities of
up to $N_{\rm H}$ of $10^{22.5}$ cm$^{-2}$ can be detected without bias
using 2--10~keV emission, the fraction of absorbed narrow-line AGNs is
consistent with the absorption column density distribution of nearby 
Seyfert galaxies (Risaliti et al. 1999).
On the other hand, in the luminosity range above $10^{45}$ erg s$^{-1}$,
14 broad-line AGNs are detected in total, but no narrow-line AGN is found.
The result
suggests lack of absorbed luminous AGNs (deficiency of type 2 QSOs).
Similar tendencies also appear in the {\it HEAO1} A2 sample 
(small circles in Figure 1; Piccinotti et al. 1982)
and recent results of {\it Chandra} optical identification
(small triangles in Figure 1; Hornschemeier et al. 2000;
Mushotzky et al. 2000; Brandt et al. 2000; Fiore et al. 2000).
{\it Chandra} AGNs with a strong broad-line(s) in their optical spectra or listed as a QSO are
plotted with open symbols, and filled symbols represent other AGNs or galaxies.
It should be noted that the optical identificaiton of {\it Chandra}
sources is not complete at this moment and unidentified X-ray sources
with red optical couterparts may be a missing population of 
type 2 QSOs.

\section{X-ray Spectra of Hard X-ray Selected QSOs}

\begin{figure}
\plotfiddle{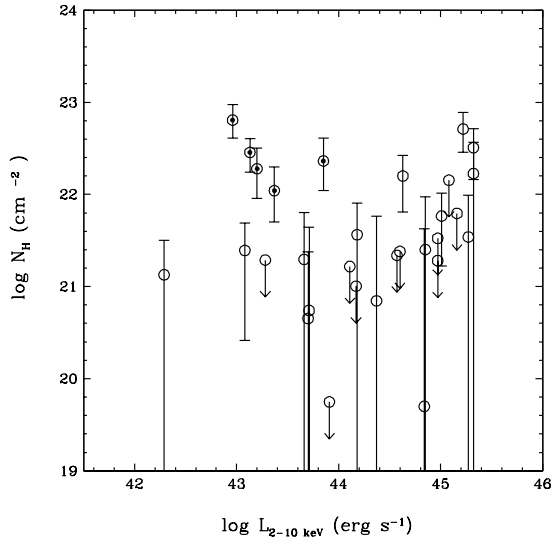}{2.7in}{0}{130}{130}{-390}{-440}
\caption{
Fitted column density versus 2--10~keV luminosity diagram of the 
{\it ASCA} LSS AGNs. The X-ray luminosities are not corrected
for the absorption. Narrow-line AGNs with X-ray absorption
with hydrogen column density larger than $10^{22}$ cm$^{-2}$ are
marked with dots. Upper limits for column densities are
indicated with downward arrows.}
\end{figure}

Some of broad-line z$\sim$1 QSOs have harder X-ray spectra
than typical X-ray spectra of QSOs (power-law with photon index
of 1.7), although their
optical colors, optical spectra, and X-ray to optical flux ratios 
are same as normal QSOs. In the {\it ASCA} LSS AGNs,
four high-redshift broad-line QSOs show hard X-ray spectra
with apparent photon indices of 1.3$\pm$0.3. The hard X-ray spectra
may also be explained by absorption with log $N_{\rm H}$(cm$^{-2}$)=22$\sim$23
at the object's redshift, if we assume an intrinsic photon index of 1.7
(Figure 2, it should be noted that redshifts and hard X-ray luminosities
correlate each other in the {\it ASCA} LSS sample. Therefore, the hard X-ray
luminosity of $L_{\rm 2--10~keV}
\sim10^{45}$ erg s$^{-1}$ corresponds to the redshift of $z\sim1$). 
Similar hardenings of X-ray spectra of
high-redshift broad-line QSOs are also reported in
other {\it ASCA} surveys (Della Ceca et al. 2000) 
and {\it Beppo-SAX} surveys (La Franca et al. 2000).
The origin of the hardness is not clear yet. The hardness of the 
X-ray spectra of high-redshift QSOs may be explained by the 
existence of the reflection component. Because of a bump of the reflection
component, observed photon indices of type 1 Seyferts in the 0.7--10~keV
band is expected to get harder to $z\sim2$.

\section{A Case of an Absorbed QSO at z=0.65}

A few candidates of narrow-line-strong absorbed QSOs 
were found in the {\it ASCA} surveys, though the fraction of absorbed
AGNs in the high luminosity range is not as high as that in the low luminosity range.
AX~J131831+3341 is a moderately absorbed ($N_{\rm H}$ of $10^{22}$ cm$^{-2}$)
 QSO at z=0.65 
found in the {\it ASCA} LSS (Akiyama et al. 2000b, Akiyama et al. 2001).
Its optical spectrum shows strong emission lines, such as
broad Mg {\scriptsize II} 2800 {\AA}, narrow
[O {\scriptsize II}] 3727 {\AA}, and narrow
[O {\scriptsize III}] 5007 {\AA}, but
no broad H$\beta$ emission line.
Its small H$\beta$-to-[O {\scriptsize III}] 5007 {\AA} equivalent
width ratio [log(H$\beta$/[O {\scriptsize III}])$=-0.54$]
is comparable to those of Seyfert 1.8--2 galaxies (Winkler 1992).
The X-ray luminosity is estimated to be $10^{45} \ {\rm erg} \ {\rm s}^{-1}$,
which is as large as the luminosity of the knee of the
AGN luminosity function in the 2--10~keV band at $z \sim 0.6$
(e.g., Boyle et al. 1998),
and corresponds to the luminosities of QSOs.

The optical and near-infrared images show a nucleus and an extended host galaxy
around the QSO (left panel in Figure 3).
In the right panel of Figure 3, the $R-I$ and $I-K$ colors of 
nuclear and host galaxy components are shown.
The nuclear component have blue $R-I$ colors but red $I-K$ color. 
The $I-K$ color of the nuclear component is
much redder than optically-selected QSOs at similar redshifts
(open squares; Elvis et al. 1994).
A possible explanation of these colors is that,
in the $K$ band, heavily absorbed ($A_V \sim 3$ mag) nucleus emerges, 
while optical light originates from scattered blue nuclear light.
The estimated fraction of scattered light is 2\%, which is similar
to that observed in narrow-line radio galaxies (Alighieri et al. 1994).
Such a red optical to near-infrared color and a blue optical color
may be common characteristics of X-ray selected absorbed QSOs 
at intermediate redshifts. Recent observations reveal that another
absorbed QSO at a redshift of 0.9, AX~J08494+4454, also has a
blue optical color and a red optical to near-infrared color 
($R-I=0.67$ mag and $I-K=3.4$ mag; Nakanishi et al. 2000; Akiyama et al. 
in preparation).
Also, a significant fraction of the optical and near-infrared counterparts
of {\it Chandra} hard X-ray sources show similar red near-infrared colors
($I-HK'=4\sim5$ mag) and blue optical colors ($B-I=1\sim2$ mag) 
(Mushotzky et al. 2000).

The host galaxy extends 62 kpc away from the nucleus.
The asymmetric and extended structure of the component suggests a galaxy
interaction or merging in the host galaxy. 
The $V-R$, $R-I$, and $I-K$ colors of the host galaxy are consistent with a stellar
population with age of 1 Gyr. Thus, AX~J131831+3341 may have a post-starburst
galaxy as the host galaxy like a "post-starburst quasar", 
UN~J1025-0040 (Brotherton et al. 1999).

\begin{figure}
\plotfiddle{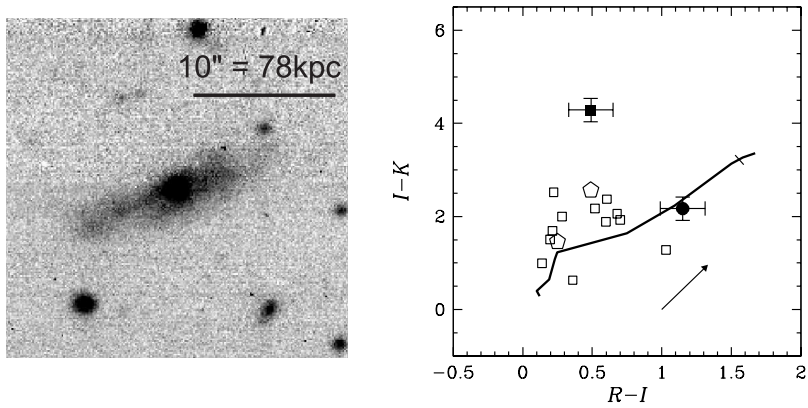}{2.5in}{0}{130}{130}{-430}{-450}
\caption{
Right) 
Optical $R$ band image of AX~J131831+3341 taken with
the 8.2m Subaru telescope with integration time of 1800 s.
Left) 
The $R-I$ and $I-K$ color-color diagram of the nuclear (filled square)
and the host galaxy (filled circle) components of AX~J131831+3341.
The open squares represents colors of optically-selected
QSOs at redshifts between 0.2 and 1.5 from Elvis et al.\ (1994).
Pentagons show colors of power-law model with
indices ($f_{\nu}=\nu^{\alpha}$)
of $-1.0$ (top) and $0.0$ (bottom).
The tracks of passively evolving stellar population models
with ages from 0.01 Gyr to 12 Gyr are indicated with
thick solid line. 
The arrow represents the effect of reddening with $A_V$ of 1 mag.
}
\end{figure}

\acknowledgments
I would like to thank members of {\it ASCA} Large Sky Survey,
and {\it ASCA} Medium Sensitivity Surveys, especially Yoshihiro Ueda,
Tadayuki Takahashi, Kouji Ohta, and Toru Yamada, and Suprime-Cam team.

\end{document}